\font\bbb=msbm10                                                   %%%
%\def\bbb{\bf}                                                     %%%
%%%                                                                %%%
%%%                                                                %%%
%%%%%%%%%%%%%%%%%%%%%%%%%%%%%%%%%%%%%%%%%%%%%%%%%%%%%%%%%%%%%%%%%%%%%%
%%%%%%%%%%%%%%%%%%%%%%%%%%%%%%%%%%%%%%%%%%%%%%%%%%%%%%%%%%%%%%%%%%%%%%

\def\N{\hbox{\bbb N}}

\def\Z{\hbox{\bbb Z}}

\def\CACM{{\sl Commun.\ ACM}}
\def\CMP{{\sl Commun.\ Math.\ Phys.}}

\def\CS{{\sl Complex Systems}}

\def\FP{{\sl Found.\ Phys.}}

\def\IJMPB{{\sl Int.\ J. Mod.\ Phys.\ B}}
\def\IJTP{{\sl Int.\ J. Theor.\ Phys.}}

\def\JAP{{\sl J. Appl.\ Phys.}}

\def\JMP{{\sl J. Math.\ Phys.}}
\def\JPA{{\sl J. Phys.\ A:  Math.\ Gen.}}

\def\JSP{{\sl J. Stat.\ Phys.}}
\def\NPB{{\sl Nucl.\ Phys.\ B}}

\def\PD{{\sl Physica D}}

\def\PLB{{\sl Phys.\ Lett.\ B}}

\def\PRA{{\sl Phys.\ Rev.\ A}}
\def\PRB{{\sl Phys.\ Rev.\ B}}
\def\PRD{{\sl Phys.\ Rev.\ D}}
\def\PRL{{\sl Phys.\ Rev.\ Lett.}}
\def\PRSLA{{\sl Proc.\ Roy.\ Soc.\ Lond.\ A}}
\def\PTP{{\sl Prog.\ Theor.\ Phys.}}
\def\PTRSLA{{\sl Phil.\ Trans.\ Roy.\ Soc.\ Lond.\ A}}
\def\Sc{{\sl Science}}
\def\SPJETP{{\sl Sov.\ Phys.\ JETP}}
\def\TIEICEJE{{\sl Trans.\ IEICE Japan E}}

\def\dajm{\hbox{D. A. Meyer}}

\def\brosl{\hbox{B. Hasslacher}}
\def\bd{\hbox{\brosl\ and \dajm}}

\def\feynman{\hbox{R. P. Feynman}}
\def\deutsch{\hbox{D. Deutsch}}
\def\teich{\hbox{W. G. Teich}}
\def\gz{\hbox{G. Gr\"ossing and A. Zeilinger}}
\def\hpdp{\hbox{J. Hardy, Y. Pomeau and O. de Pazzis}}
\def\hdpp{\hbox{J. Hardy, O. de Pazzis and Y. Pomeau}}

\catcode`@=11
\newskip\ttglue

   \font\ninerm=cmr9    \font\eightrm=cmr8   \font\sixrm=cmr6
  \font\ninebf=cmbx9   \font\eightbf=cmbx8  \font\sixbf=cmbx6
  \font\nineit=cmti9   \font\eightit=cmti8  
  \font\ninesl=cmsl9   \font\eightsl=cmsl8  
  \font\ninemi=cmmi9   \font\eightmi=cmmi8  \font\sixmi=cmmi6

\font\bigtenbf=cmr10 scaled\magstep2 

\def\ninepoint{\def\rm{\fam0\ninerm}%
  \textfont0=\ninerm \scriptfont0=\sixrm
  \textfont1=\ninemi \scriptfont1=\sixmi
  \textfont\itfam=\nineit  \def\it{\fam\itfam\nineit}%
  \textfont\slfam=\ninesl  \def\sl{\fam\slfam\ninesl}%
  \textfont\bffam=\ninebf  \scriptfont\bffam=\sixbf
    \def\bf{\fam\bffam\ninebf}%
  \tt \ttglue=.5em plus.25em minus.15em
  \normalbaselineskip=11pt
  \setbox\strutbox=\hbox{\vrule height8pt depth3pt width0pt}%
  \normalbaselines\rm}

\def\eightpoint{\def\rm{\fam0\eightrm}%
  \textfont0=\eightrm \scriptfont0=\sixrm
  \textfont1=\eightmi \scriptfont1=\sixmi
  \textfont\itfam=\eightit  \def\it{\fam\itfam\eightit}%
  \textfont\slfam=\eightsl  \def\sl{\fam\slfam\eightsl}%
  \textfont\bffam=\eightbf  \scriptfont\bffam=\sixbf
    \def\bf{\fam\bffam\eightbf}%
  \tt \ttglue=.5em plus.25em minus.15em
  \normalbaselineskip=9pt
  \setbox\strutbox=\hbox{\vrule height7pt depth2pt width0pt}%
  \normalbaselines\rm}

\def\sfootnote#1{\edef\@sf{\spacefactor\the\spacefactor}#1\@sf
      \insert\footins\bgroup\eightpoint
      \interlinepenalty100 \let\par=\endgraf
        \leftskip=0pt \rightskip=0pt
        \splittopskip=10pt plus 1pt minus 1pt \floatingpenalty=20000
        \parskip=0pt\smallskip\item{#1}\bgroup\strut\aftergroup\@foot\let\next}
\skip\footins=12pt plus 2pt minus 2pt
\dimen\footins=30pc

\def\ie{{\it i.e.}}
\def\eg{{\it e.g.}}

\def\etal{{\it et al.}}

\def\Lemma{L{\eightpoint EMMA}}

\def\endproof{\vrule height6pt width6pt depth0pt}

\def\hfb{\hfil\break}

\magnification=1200
\input epsf.tex

\dimen0=\hsize \divide\dimen0 by 13 \dimendef\chasm=0
\dimen1=\hsize \advance\dimen1 by -\chasm \dimendef\usewidth=1
\dimen2=\usewidth \divide\dimen2 by 2 \dimendef\halfwidth=2
\dimen3=\usewidth \divide\dimen3 by 3 \dimendef\thirdwidth=3
\dimen4=\hsize \advance\dimen4 by -\halfwidth \dimendef\secondstart=4
\dimen5=\halfwidth \advance\dimen5 by -\parindent \dimendef\indenthalfwidth=5

\parskip=0pt
\line{\hfil August 1995}
\line{\hfil{\it revised\/} March 1996}
\line{\hfil quant-ph/9604003}
\bigskip\bigskip\bigskip
\centerline{\bf\bigtenbf FROM QUANTUM CELLULAR AUTOMATA}
\bigskip
\centerline{\bf\bigtenbf TO QUANTUM LATTICE GASES}
\vfill
\centerline{\bf David A. Meyer}
\bigskip 
\centerline{\sl Project in Geometry and Physics}
\centerline{\sl Department of Mathematics}
\centerline{\sl University of California/San Diego}
\centerline{\sl La Jolla, CA 92093-0112}
\centerline{dmeyer@euclid.ucsd.edu}
\vfill
\centerline{ABSTRACT}
\bigskip
%--------|---------|---------|---------|---------|---------|---------|
\noindent A natural architecture for nanoscale quantum computation is 
that of a quantum cellular automaton.  Motivated by this observation,
in this paper we begin an investigation of exactly unitary cellular 
automata.  After proving that there can be no nontrivial, homogeneous, 
local, unitary, scalar cellular automaton in one dimension, we weaken 
the homogeneity condition and show that there are nontrivial, exactly
unitary, partitioning cellular automata.  We find a one parameter 
family of evolution rules which are best interpreted as those for a 
one particle quantum automaton.  This model is naturally reformulated
as a two component cellular automaton which we demonstrate to limit to 
the Dirac equation.  We describe two generalizations of this 
automaton, the second of which, to multiple interacting particles, is 
the correct definition of a quantum lattice gas.
\bigskip
%--------|---------|---------|---------|---------|---------|---------|

\vfill
\centerline{\JSP\ {\bf 85} (1996) 551--574.}
\vfill
\eject

\headline{\ninepoint\it From QCA to QLGA\hfil David A. Meyer}
\parskip=10pt

\noindent{\bf 1.  Introduction}

%--------|---------|---------|---------|---------|---------|---------|
\noindent The realization that incentives to develop smaller and 
faster computers will eventually drive the devices from which they are 
constructed into the quantum regime motivated research into quantum
mechanical limitations on deterministic computation as early as the 
1970s [1].  The subsequent conceptions of universal quantum simulator 
by Feynman [2] and quantum Turing machine by Deutsch [3] initiated a 
series of investigations [4] into how aspects of quantum mechanics, 
specifically superposition and interference, might be exploited for 
computational purposes.  Shor's remarkable discovery of a polynomial 
time quantum algorithm for factorization [5] (and the centrality of 
the factoring problem to modern cryptography [6]) has led to redoubled 
interest in the design and construction of quantum computational 
nanodevices [7].  It should be emphasized that the goal here is a 
computational device which will run {\sl quantum\/} algorithms, not a 
quantum device which will run deterministic [8,9] or probabilistic 
[10] algorithms.

%--------|---------|---------|---------|---------|---------|---------|
For a variety of reasons---the wire and gain problems, and the 
pragmatic observation that an array of simple devices is often easier 
to design and build than a single, more complicated device---it seems 
likely that massive parallelism will optimize nanoscale computer 
architecture [11].  In this paradigm, a quantum computer is a quantum 
cellular automaton (QCA):  the state of each simple device (cell) in 
the array depends on the states of the cells in some local 
neighborhood at the previous timestep.  Unlike the original cellular 
automaton (CA) models for parallel computation of von Neumann and 
Ulam [12], where this dependence is deterministic or probabilistic, 
here the dependence is quantum mechanical:  There is a (complex) 
probability amplitude for the transition to each possible state, 
subject to the condition that the evolution be unitary, so that the 
total probability---the sum of the norm squared of the amplitude of 
each configuration---is always one.

%--------|---------|---------|---------|---------|---------|---------|
QCA, therefore, provide a laboratory for analyzing both potential 
quantum computer architectures and algorithms; this is the motivation 
for initiating our study of them in this paper.  Computation motivated 
study of QCA seems to have originated with the interesting work of 
Gr\"ossing and Zeilinger [13].  Their models, however, are only 
approximately quantum mechanical because, they argue, ``except for the 
trivial case, strictly local, unitary evolution {\sl of the whole QCA 
array\/} is impossible'' [14].  Consequently, they study CA whose 
evolution is both nonunitary [13] and, although `probability' 
preserving, nonlinear [14].  In Section 2 we begin by proving the
following precise version of their claim:

%--------|---------|---------|---------|---------|---------|---------|
\noindent N{\eightpoint O-GO} \Lemma.  {\sl In one dimension there 
exists no nontrivial, homogeneous, local, scalar QCA.  More 
explicitly, every band $r$-diagonal unitary matrix which commutes with 
the 1-step translation matrix is also a translation matrix, times a 
phase.}

%--------|---------|---------|---------|---------|---------|---------|
\noindent and then continue by showing that even a slight weakening of 
the homogeneity/translation invariance condition allows nontrivial 
unitary evolution.  Thus we will reserve the adjective `quantum' for
CA with exactly unitary, nontrivial, local evolution, in contrast to 
the usage by Gr\"ossing, Zeilinger, \etal\ [13,14] and Lent and Tougaw 
[9].  These are the QCA which should best model truly quantum 
parallel architectures.  In this paper we consider only the one 
dimensional situation; but this is not irrelevant for computational 
complexity issues since reversible CA capable of universal 
computation exist in one dimension [15], and reversible deterministic 
CA are, of course, unitary.

%--------|---------|---------|---------|---------|---------|---------|
Section 3 contains output from several simulations of one dimensional 
QCA.  Simulation on a deterministic computer must inevitably be slow, 
but this is acceptable since our goal here is understanding rather 
than the solution of any specific problem.  It is easy to see 
qualitative differences from the simulations of Gr\"ossing, Zeilinger, 
\etal\ [13,14]; in particular, our simulations display particle-like 
features.

%--------|---------|---------|---------|---------|---------|---------|
The latter observation motivates our reinterpretation in Section 4 of
the evolution rule of such a QCA as the scattering rule for a quantum 
particle automaton.  In this form the model is equivalent to Feynman's 
path integral formulation for a Dirac particle [16]; it is 
straightforward to solve the model exactly and to give a physical 
interpretation to the parameter in the scattering rule.

%--------|---------|---------|---------|---------|---------|---------|
A lattice gas [17] formulation of this quantum particle automaton 
would consist of an array of nodes occupied by left and/or right 
moving particles which jump to the next node at the next timestep.  
Since there is only {\sl one\/} particle in the model the amplitudes 
for left and right moving particles at a given node may be combined 
into a two component field as in the one dimensional Dirac equation in 
the chiral representation.  Formulated as such a two component QCA, 
our quantum particle automaton is unitarily equivalent to the quantum 
lattice Boltzmann equation of Succi and Benzi [18] and to the one 
dimensional version of Bialynicki-Birula's QCA [19].  Since a two 
component QCA evades the conclusion of the No-go Lemma, in Section 5 
we find the most general homogeneous/translation invariant unitary 
evolution rules for a neighborhood of radius one.  The lattice gas 
paradigm motivates another generalization, however, to multiple 
(interacting) quantum particles.  This new system, described in 
Section 6, is a quantum lattice gas automaton (QLGA) and may be 
expected to be relevant for modelling parallel quantum architectures 
constructed from few electron devices.

%--------|---------|---------|---------|---------|---------|---------|
As Landauer [20] and others [21] have emphasized, achieving practical 
quantum computation will be difficult for a variety of reasons.  The 
most fundamental problem is decoherence [22,23], which destroys the 
delicate interference phenomena on which quantum algorithms such as 
Shor's [5] depend.  In Section 7 we observe that this is among the 
important issues in quantum computation which can be investigated 
using these models and also remark on some related research 
directions.

\medskip
\noindent{\bf 2.  Quantum cellular automata}
\nobreak

\nobreak
%--------|---------|---------|---------|---------|---------|---------|
\noindent A CA consists of a lattice $L$ of {\sl cells\/} together 
with a {\sl field\/} $\phi : \N \times L \to S$, where $\N$ denotes 
the non-negative integers labelling timesteps and $S$ is the set of 
possible {\sl states\/} in which the field is valued.  The field 
evolves according to some {\sl local rule\/}, \ie, $\phi$ satisfies a 
recurrence relation of the form
$$
\phi_{t+1}(x) = f\bigl(\phi_t(x+e) \mid e \in E(t,x)\bigr),   \eqno(1)
$$
where $E(t,x)$ is a set of lattice vectors defining local 
{\sl neighborhoods\/} for the automaton.

%--------|---------|---------|---------|---------|---------|---------|
In the Schr\"odinger picture of quantum mechanics the state of a 
system at time $t$ is a {\sl state vector\/} in some Hilbert space
(see, \eg, [24]).  The state vector evolves locally and unitarily, 
\ie, 
$$
\phi_{t+1} = U\phi_t,                                         \eqno(2)
$$
where $U$ is a {\sl unitary\/} matrix (more precisely, operator:  
$UU^{\dagger} = I = U^{\dagger}U$).  Thus, if the configuration space 
is discrete, and the Hilbert space has a computational basis  
$|x\rangle$, $x \in L$, it is natural to try to construct a QCA model 
where $\phi_t(x)$ is the (complex scalar) coefficient of $|x\rangle$ 
in $\phi_t$ and the local evolution rule (1) is the unitary evolution 
rule (2) in this basis.  Notice that this forces the QCA to be 
{\sl additive\/} in Wolfram's terminology [25], \ie, (1) becomes
$$
\phi_{t+1}(x) = \sum_{e\in E(t,x)} w(t,x+e) \phi_t(x+e),      \eqno(3)
$$
where the coefficients $w(t,x+e)$ are constrained by the unitarity 
condition.  If both $E(t,x)$ and $w(t,x+e)$ are independent of $t$ and
$x$, the CA is {\sl homogeneous}.

%--------|---------|---------|---------|---------|---------|---------|
In one dimension $\phi_t$ can be written as a column vector with 
ordered entries $\phi_t(x)$, in which case locality corresponds to $U$
being band diagonal.  If, furthermore, the QCA is homogeneous, $U$ 
must be invariant under the action of the translation operator $T$ on 
the lattice---the permutation matrix with 1s on the subdiagonal---and 
(2) becomes
$$
\pmatrix{     \vdots    \cr
         \phi_{t+1}(-1) \cr
         \phi_{t+1}(0)  \cr
         \phi_{t+1}(+1) \cr
              \vdots    \cr
        }
= 
\pmatrix{
\ddots&        &        &        &        &        &      \cr
      & w_{-r} & \ldots & w_{+r} &        &        &      \cr
      &        & w_{-r} & \ldots & w_{+r} &        &      \cr
      &        &        & w_{-r} & \ldots & w_{+r} &      \cr
      &        &        &        &        &        &\ddots\cr
}
\pmatrix{  \vdots   \cr
         \phi_t(-1) \cr
         \phi_t(0)  \cr
         \phi_t(+1) \cr
           \vdots   \cr
        },                                                    \eqno(4)
$$
where $w(t,x+e) \equiv w_e$.  (With periodic boundary conditions, of 
course, the top and bottom $r$ rows will wrap around.)  Gr\"ossing and 
Zeilinger consider the case $r = 1$ but find no nontrivial unitary 
matrix $U$.  More generally,

%--------|---------|---------|---------|---------|---------|---------|
\noindent N{\eightpoint O-GO} \Lemma.  {\sl In one dimension there 
exists no nontrivial, homogeneous, local, scalar QCA.  More 
explicitly, every band $r$-diagonal unitary matrix\/} $U$ {\sl which 
commutes with the 1-step translation matrix\/} $T$ {\sl is also a 
translation matrix\/} $T^k$ {\sl for some\/} $k \in \Z$, {\sl times a 
phase.}

%--------|---------|---------|---------|---------|---------|---------|
\noindent{\sl Proof}.  When $r = 0$, $U = w_0 I$ and unitarity implies
$|w_0|^2 = 1$ so $U$ is the 0-step translation matrix (the identity) 
times a phase.  Assume that the statement is true for $r-1$.  For a 
1-step translation invariant band $r$-diagonal matrix, unitarity 
implies:
\vbox{
$$
\setbox2=\hbox{${}={}$}
\def\svdots{\hbox to\wd2{\hfil\vdots\hfil}}
\openup1\jot \tabskip=0pt plus1fil
\halign to \displaywidth{\tabskip=0pt
$#\hfil$&$\hfil{}#{}$&
$#\hfil$&$\hfil{}#{}$&
$#\hfil$&$\hfil{}#{}$&
$#\hfil$&$\hfil{}#{}$&
$#\hfil$&${}#\hfil$\tabskip=0pt plus1fil&\llap{$#$}\tabskip=0pt\cr
w_{-r} \bar w_{-r}&+&
w_{-r+1} \bar w_{-r+1}&+&
\cdots&+&
w_{r-1} \bar w_{r-1}&+&
w_r \bar w_r&=1&(5_{-r})\cr
&&
w_{-r+1} \bar w_{-r}&+&
\cdots&+&
w_{r-1} \bar w_{r-2}&+&
w_r \bar w_{r-1}&=0&(5_{-r+1})\cr
&&&&&&&&&\svdots&\cr
&&&&&&
w_{r-1} \bar w_{-r}&+&
w_r \bar w_{-r+1}&=0&(5_{r-1})\cr
&&&&&&&&
w_r \bar w_{-r}&=0,&(5_r)\cr
}
$$
}
\noindent together with the complex conjugate equations.  By equation 
$(5_r)$, at least one of $w_r$ and $w_{-r}$ vanishes, say $w_r$.  Then 
we may assume $w_{-r} \not= 0$; otherwise the conclusion follows 
immediately from the inductive hypothesis.  But then equation 
$(5_{r-1})$ forces $w_{r-1} = 0$; $\dots$; equation $(5_{-r+1})$ 
forces $w_{-r+1} = 0$; and equation $(5_{-r})$ becomes 
$|w_{-r}|^2 = 1$, \ie, the only nonzero weight, $w_{-r}$, is a phase.  
Then $U = w_{-r} T^r$, where $T^r$ is the $r$-step translation matrix.
                                                       \hfill\endproof

%--------|---------|---------|---------|---------|---------|---------|
CA which evolve simply by translation are not very interesting, so 
Gr\"ossing and Zeilinger relax the unitarity constraint, setting
$$
w_{-1} = i\delta, \quad
 w_{0} = 1, \quad
w_{+1} = i\bar\delta,
$$
so that the evolution is only approximately unitary, with errors of 
$O(|\delta|^2)$ accumulating at each timestep [13].  Since the 
evolution is nonunitary, {\sl quantum\/} probability is not preserved.
Instead, all the amplitudes are normalized by an overall factor at 
each step to make $\sum_x |\phi_t(x)|^2 = 1$; this may be thought of 
as a nonlocal (and non-quantum mechanical) aspect of the evolution 
[14].

%--------|---------|---------|---------|---------|---------|---------|
Our choice, instead, is to weaken the homogeneity condition but to 
insist still on exactly unitary, local evolution, thus maintaining
consistency with standard quantum mechanics.  The evolution matrix in 
(4) is 1-step translation invariant, \ie, $T U T^{-1} = U$; a natural 
way to relax this constraint is to require $T^2 U T^{-2} = U$, so that 
the evolution is only 2-step translation invariant.  For $r = 1$ we 
have then:
$$
U = \pmatrix{\ddots &   &   &   &   &   &   &      \cr
                    & a & b & c &   &   &   &      \cr
                    &   & d & e & f &   &   &      \cr
                    &   &   & a & b & c &   &      \cr
                    &   &   &   & d & e & f &      \cr
                    &   &   &   &   &   &   &\ddots\cr
            },
$$
and unitarity requires:
$$
\eqalign{a \bar a + b \bar b + c \bar c &= 1  \cr
                    b \bar d + c \bar e &= 0  \cr
                               c \bar a &= 0  \cr
        }
\qquad
\eqalign{d \bar d + e \bar e + f \bar f &= 1  \cr
                    e \bar a + f \bar b &= 0  \cr
                               f \bar d &= 0, \cr
        }                                                     \eqno(6)
$$
together with the complex conjugate equations.

%--------|---------|---------|---------|---------|---------|---------|
There are two types of solutions to these equations.  The 
uninteresting ones have only $a$ and $d$ (or $c$ and $f$) nonzero, 
both with norm 1; in this case the evolution is by translation and
multiplication by alternating phases.  The interesting solutions to
(6) have $c = d = 0$ (or $a = f = 0$) and the matrix 
$$
S := \pmatrix{e & f \cr a & b \cr} \quad
\hbox{(or $\pmatrix{ b & c \cr d & e \cr}$)}                  \eqno(7)
$$
unitary; in this case $U$ is block diagonal, each block acting only on
a pair of adjacent cells.

%--------|---------|---------|---------|---------|---------|---------|
Evolving by $U$ at each timestep would partition the CA into a set of 
noninteracting systems each comprised of a pair of adjacent cells.  
Instead, we evolve by $U$ and by $T U T^{-1}$ at alternating 
timesteps, changing the pairing and allowing propagation.  Such 
alternating evolution has been referred to as a {\sl staggered\/} rule 
for a {\sl checkerboard\/} model in the context of probabilistic CA
and two dimensional statistical mechanics models [26], and as a 
{\sl partitioning\/} CA in the context of reversible CA [27].  Since 
unitary evolution includes deterministic reversible evolution, the
No-go Lemma applies in the latter context in one dimension (of course, 
the overall phase referred to in the statement of the lemma must be 
1), and it is natural that triviality has been evaded there in the 
same way.

\medskip
\noindent{\bf 3.  Simulations}
\nobreak

\nobreak
%--------|---------|---------|---------|---------|---------|---------|
\noindent The matrix $S \in U(2)$ may be parameterized as
$$
S = \pmatrix{e^{i\alpha}\sin\theta  & e^{i\beta} \cos\theta  \cr
             e^{i\gamma}\cos\theta  & e^{i\delta}\sin\theta  \cr
            },
$$
with $\alpha - \beta - \gamma + \delta \equiv \pi$ (mod $2\pi$).  
Rather than continuing in full generality, we shall impose parity
invariance, which forces $b = e$ and $a = f$ (or $b = e$ and $c = d$)
in (7).  Dividing out an overall phase which is unobservable (\ie, 
has no effect on probabilities), the cell pair evolution matrix $S$
becomes
$$
S = \pmatrix{i\sin\theta &  \cos\theta \cr
              \cos\theta & i\sin\theta \cr
            }.                                                \eqno(8)
$$

%--------|---------|---------|---------|---------|---------|---------|
Simulation of a single timestep of this QCA is achieved by a series of
matrix multiplications:
$$
\pmatrix{\phi_{t+1}(x-1)  \cr
         \phi_{t+1}(x)    \cr
        }
=
\pmatrix{i\sin\theta &  \cos\theta \cr
          \cos\theta & i\sin\theta \cr
        }
\pmatrix{\phi_t(x-1)  \cr
         \phi_t(x)    \cr
        },                                                    \eqno(9)
$$
for each $x \equiv t+1$ (mod 2), say.  Although these matrix 
multiplications commute since they evolve disjoint pairs of cells, 
sequential simulation of this intrinsically parallel evolution will be 
slow, as it always is for CA.  Furthermore, since the field values 
and the evolution parameters are complex numbers, the usual
efficiencies of bitwise computation are unavailable.  Nevertheless,
simulation of small systems is easy and informative; an additional,
more serious difficulty will not appear until the simulations of QLGA 
in Section 5.

%--------|---------|---------|---------|---------|---------|---------|
Figure 1 shows simulations of our QCA for $\theta = \pi/6$ and 
$\theta = \pi/3$, starting from the same random initial condition.  
The darkness of each cell is (positively) proportional to its 
probability, \ie, the norm squared of its amplitude (field value), and 
it appears that there is a probability flow in each case, slower for 
$\theta = \pi/3$.  This is expected since for $\theta = \pi/2$, $U$ is 
proportional to the identity (the overall factor of $i$ is 
unobservable) and there is no flow, while for $\theta = 0$, $S$ simply 
interchanges the states of adjacent cells so that probability 
propagates with speed 1 in lattice units.

%--------|---------|---------|---------|---------|---------|---------|
\topinsert
$$
\epsfxsize=\thirdwidth\epsfbox{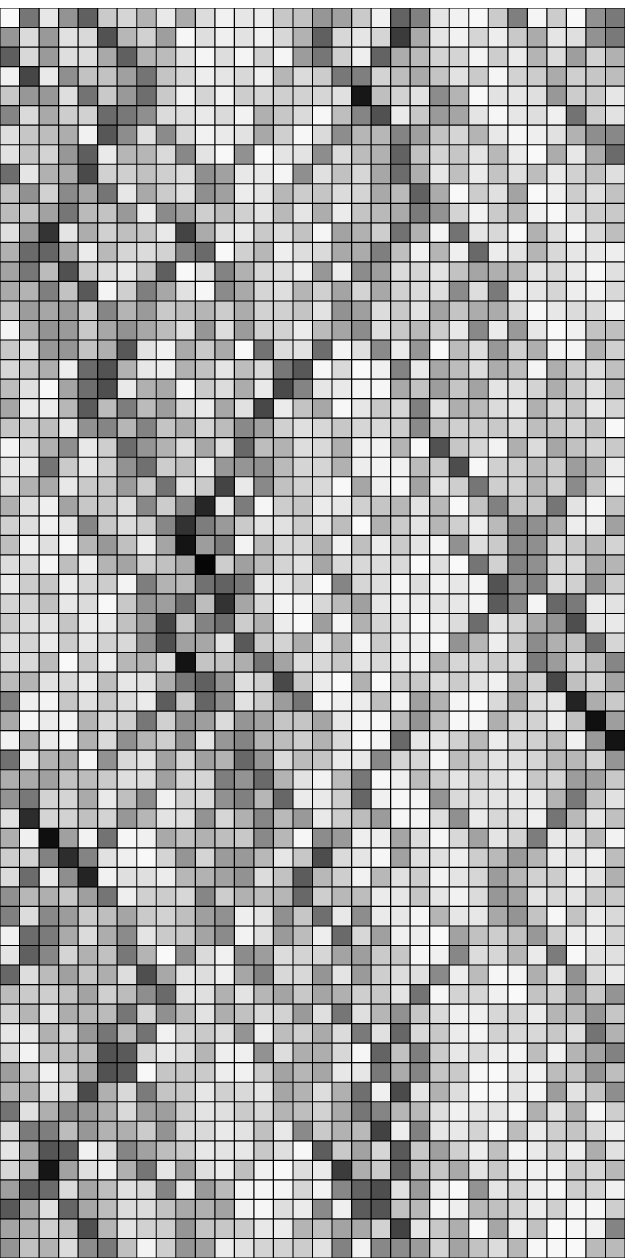}\hskip0.75truein%
\epsfxsize=\thirdwidth\epsfbox{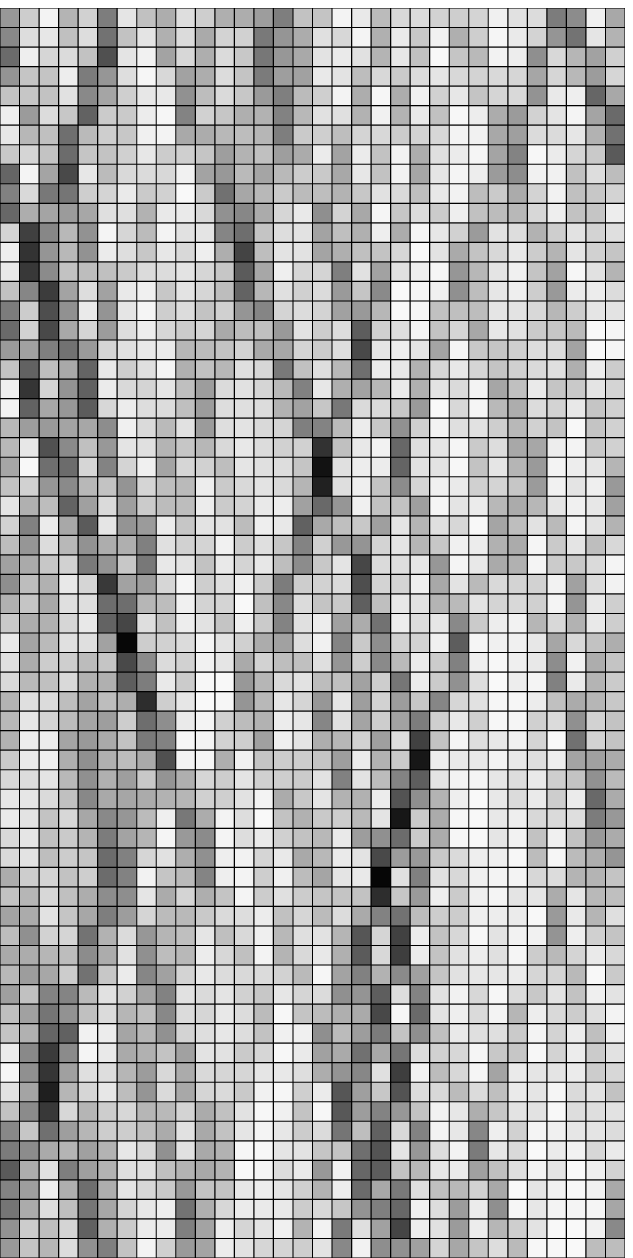}
$$
\eightpoint{%
{\narrower\noindent{\bf Figure 1}.  Two simulations of the $r = 1$ QCA 
starting with the same initial conditions.  On the left 
$\theta = \pi/6$ and on the right $\theta = \pi/3$.  Time runs upward 
and periodic boundary conditions have been imposed.\par}
}
\endinsert

%--------|---------|---------|---------|---------|---------|---------|
Starting with the simplest symmetric initial condition demonstrates 
the probability propagation more clearly.  In the simulation shown in
Figure 2, the initial condition is
$$
\phi(x) = \cases{1/\sqrt{2} & if $x \in \{15,16\}$;  \cr
                         0  & otherwise.             \cr
                }
$$
Here $\theta = \pi/4$.  Measuring the locations of the two peaks of 
the probability distribution at successive timesteps indicates that 
the propagation speed is approximately $2/3$ in lattice units.  

%--------|---------|---------|---------|---------|---------|---------|
\pageinsert
$$
\epsfxsize=\thirdwidth\epsfbox{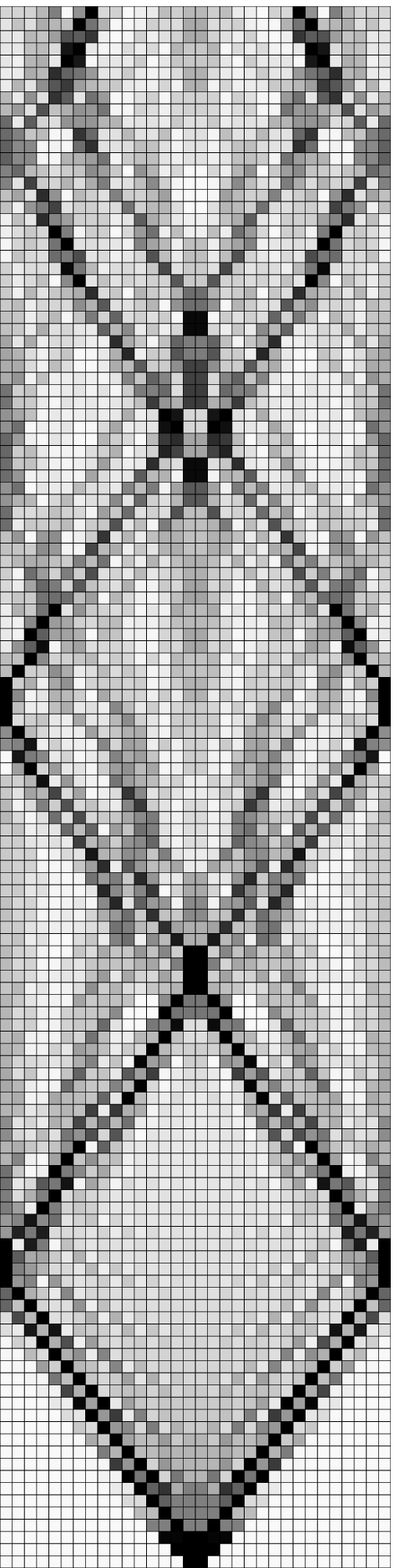}\hskip1.5truein%
\epsfxsize=\thirdwidth\epsfbox{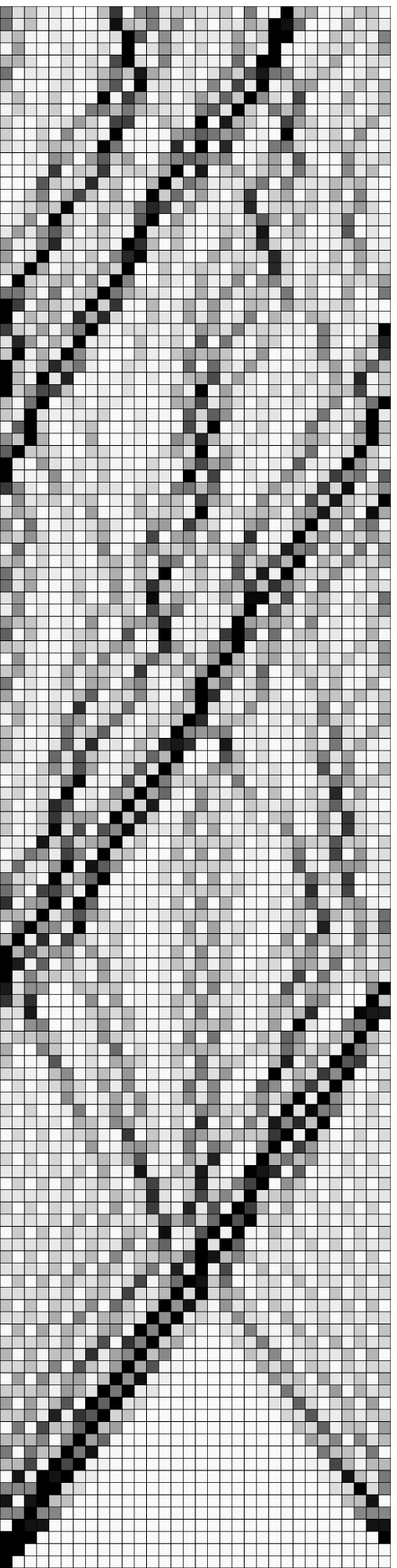}
$$
\hbox to\hsize{%
\vbox{\hsize=\halfwidth\eightpoint{%
\noindent{\bf Figure 2}.  The $r = 1$ QCA with $\theta = \pi/4$ and 
equal initial nonzero amplitudes at $x = 15,16$.
}}
\hfill%
\vbox{\hsize=\halfwidth\eightpoint{%
\noindent{\bf Figure 3}.  The same QCA with initial nonzero amplitude 
only at $x = 0$.
}}}
\endinsert

%--------|---------|---------|---------|---------|---------|---------|
This is consistent with the evolution of the even simpler initial 
condition shown in Figure 3.  Here the only nonzero initial value is 
at $x = 0$.  Again $\theta = \pi/4$ and the propagation speed still
appears to be approximately $2/3$ in lattice units.  In both of these
simulations there are peaks in the probability distribution which
remain substantially localized for the duration of the evolution
shown.  This behavior, particularly in the symmetric simulation shown
in Figure 2, should be contrasted with the results of Gr\"ossing,
Zeilinger, \etal\ [13,14], which demonstrate quite different 
qualitative features:  In their simulations macroscopic patterns 
develop and there is nothing which has the particle-like appearance of
the persistent localized peaks in these probability distributions.

\medskip
\noindent{\bf 4.  Quantum particle automata}
\nobreak

\nobreak
%--------|---------|---------|---------|---------|---------|---------|
\moveright\secondstart\vtop to 0pt{\hsize=\halfwidth
\null\vskip-\baselineskip
\vskip-\baselineskip
$$
\epsfxsize=\thirdwidth\epsfbox{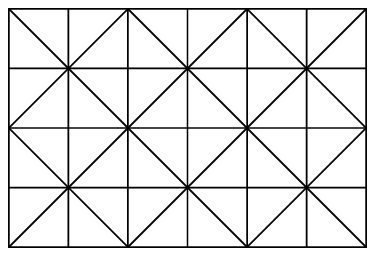}
$$
\eightpoint{%
\noindent{\bf Figure 4}.  The spacetime CA lattice is the rectangular 
array of square cells; the dual particle trajectory lattice consists 
of the diagonal edges.
}}
\vskip-\baselineskip
\parshape=11
0pt \halfwidth
0pt \halfwidth
0pt \halfwidth
0pt \halfwidth
0pt \halfwidth
0pt \halfwidth
0pt \halfwidth
0pt \halfwidth
0pt \halfwidth
0pt \halfwidth
0pt \hsize
%--------|---------|---------|---------|---------|---------|---------|
{\noindent The underlying particle nature of this 
alternating/partitioning QCA can be seen on the {\sl trajectory 
lattice\/} dual to the spacetime lattice of the CA.  As shown in 
Figure 4, each adjacent pair of cells acted on by $S$ has a dual pair 
of spacetime edges which intersect and then continue along the same 
trajectories at the next timestep, becoming the spacetime edges dual 
to the same pair of cells.  On the spacetime trajectory lattice the 
alternating action of what should now be called the {\sl scattering  
matrix\/} $S$ is automatic; the values of $\phi$ are attached to the 
edges of the trajectory lattice and undergo a unitary transformation 
by $S$ at each vertex.}

%--------|---------|---------|---------|---------|---------|---------|
The value of $\phi$ on a left/right pointing edge of the trajectory 
lattice should be interpreted as the amplitude for a left/right moving
particle being there.  That is, equation (9) means that if there is
a right moving particle at time $t$, at time $t+1$ it either becomes a 
left moving particle with amplitude $i\sin\theta$ or continues moving 
right with amplitude $\cos\theta$.  In particular, if there is some 
nonzero amplitude on each of two intersecting right and left going 
edges of the trajectory lattice, they may be evolved independently and 
then added.  There is no interaction, so this is effectively a model 
of a {\sl single\/} particle; the QCA is a {\sl quantum particle 
automaton}.  $\phi_t(x)$ is the amplitude for the particle being in 
state $|x\rangle$ during the time interval $(t,t+1)$, where, if $x$ 
takes integer values on the vertical lines in Figure 4,
$$
|x\rangle := 
\cases{\hbox{{\it left\/} moving from $x+1$} & if 
                                            $t\not\equiv x$ (mod 2)\cr 
       \hbox{{\it right\/} moving from $x$}  & if
                                            $t\equiv x$ (mod 2).   \cr
      }                                                      \eqno(10)
$$
That is, the computational basis $|x\rangle$ is a set of eigenstates 
of an operator measuring more than just position.  Projecting onto the
position subspace adds (incoherently, \ie, the {\sl probabilities\/}
add) the amplitudes for left and right movers existing at $(t,x)$, for 
$t \equiv x$ (mod 2), and gives a much clearer picture of the
evolution.  Figures 5 and 6 show this representation of the 
simulations of Figures 2 and 3, respectively.  Since the evolution 
lies within the `lightcone', \ie, the propagation speed is less that 1 
in lattice units, Figures 5 and 6 are in lightcone coordinates:
$$
\eqalign{
u &:= \bigl(t + (x - x_0)\bigr)/2                                  \cr
v &:= \bigl(t - (x - x_0)\bigr)/2,                                 \cr
}
$$
where $x_0$ implements a translation of the origin:  by 16 in Figure 5 
and by 0 in Figure 6.  In both figures only the spacetime region 
covered by $0 \le u,v \le 16$ is shown, the origin is at the rear 
corner, and time runs forward to the front corner.  The parity 
dependence (visible in the patterns of alternating dark and light 
cells) in Figures 2 and 3 has been smoothed out in these so that both 
the probable particle trajectories as well as their wave-like 
character are clearly visible.

%--------|---------|---------|---------|---------|---------|---------|
\topinsert
\vskip-\chasm
\vskip-\baselineskip
\vskip-\baselineskip
$$
\epsfxsize=\halfwidth\epsfbox{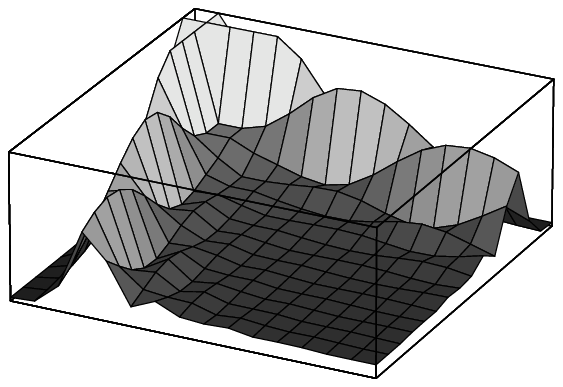}\hskip\chasm%
\epsfxsize=\halfwidth\epsfbox{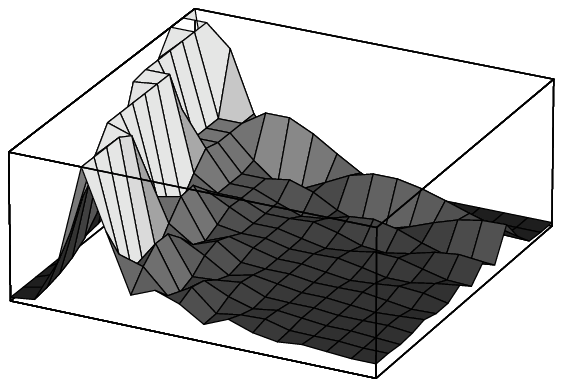}
$$
\vskip-\chasm
\hbox to\hsize{%
\vbox{\hsize=\halfwidth\eightpoint{%
\noindent{\bf Figure 5}.  Position measurements in lightcone 
coordinates for the QCA of Figure 2.  The probability 1 measurement 
at $(0,0)$ has been clipped. 
}}
\hfill%
\vbox{\hsize=\halfwidth\eightpoint{%
\noindent{\bf Figure 6}.  The same representation of the QCA of 
Figure 3.  The first few (large) probabilities in the evolution have 
been clipped.
}}}
\endinsert

%--------|---------|---------|---------|---------|---------|---------|
That the solution is as smooth as shown here suggests that the quantum
particle automaton may be a discrete approximation to a continuum 
system, possibly with an exact solution.  This is, in fact, the case.  
Although we have been led to it simply by the  assumptions of 
discreteness, locality, and unitarity (and parity invariance), we will 
show that this quantum particle automaton is a discrete approximation 
to the Feynman path integral for a Dirac particle in one dimension 
[16].  Furthermore, it is exactly solvable, even without going to the 
continuum limit.

%--------|---------|---------|---------|---------|---------|---------|
\moveright\secondstart\vtop to 0pt{\hsize=\halfwidth
{}\vskip-3pt
\vskip-\baselineskip
$$
\epsfxsize=\thirdwidth\epsfbox{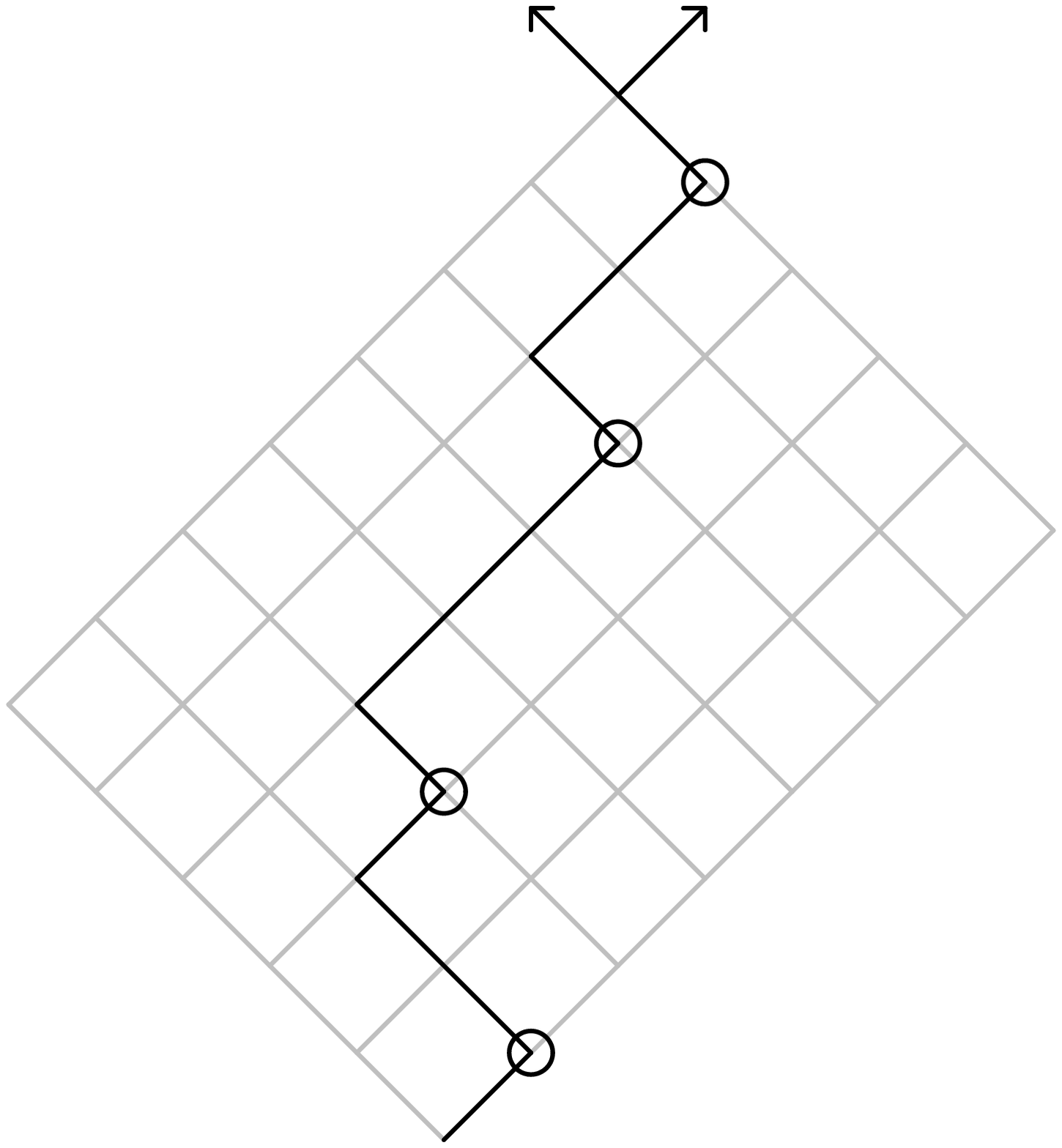}
$$
\eightpoint{%
\noindent{\bf Figure 7}.  A typical initially right moving path from 
$(u,v) = (0,0)$ to $(7,5)$ with both left and right moving final 
steps and four right to left direction changes.
}}
\vskip-\baselineskip

\parshape=2
\parindent \indenthalfwidth
0pt \halfwidth
%--------|---------|---------|---------|---------|---------|---------|
\noindent Consider a right moving particle at $(t,x) = (0,0)$, \ie, 
$\phi_0(0) = 1$.  The amplitude $\phi_t(x)$ for the particle being in 
state $|x\rangle$ at time $t$ is defined to be the {\sl propagator\/} 
$K(t,x;0,0)$.  This is exactly the (coherent) sum of the amplitudes of
all possible paths in the trajectory lattice from $(0,0)$ to 
$(t,x)$---the sum-over-histories, or Feynman path sum.  Figure 7 
shows two typical paths a right moving particle from $(0,0)$ might 
take, arriving in left moving state $|x-1\rangle$ or right moving 
state $|x\rangle$ just after time $t$ ($\equiv x$ (mod 2)).  Each path 
consists of $u$ steps right and $v$ steps left, followed by the 
terminal step, and is\break

%--------|---------|---------|---------|---------|---------|---------|
\noindent completely characterized by the locations of the right to 
left direction changes---the points circled in Figure 7.  If there are 
$k$ such direction changes, a path ending at $|x-1\rangle$ has 
amplitude
$$
\eqalignno{
(i\sin\theta)^{2k-1} (\cos\theta)^{u+v-(2k-1)} 
&=
(\cos\theta)^{u+v} (i\tan\theta)^{2k-1}                       &(11)\cr
\noalign{\smallskip\smallskip%
\hbox{and one ending at $|x\rangle$ has amplitude}%
\smallskip\smallskip}
(i\sin\theta)^{2k} (\cos\theta)^{u+v-2k}
&=
(\cos\theta)^{u+v} (i\tan\theta)^{2k}.                        &(12)\cr
}
$$
Counting the numbers of paths with $k$ direction changes and 
multiplying by the amplitudes in (11) and (12) gives
$$
\eqalignno{
\phi_t(x-1) 
&=
(\cos\theta)^{u+v}\cdot i\tan\theta \,
\sum_{k=1}^{\infty} {u-1 \choose k-1} {v \choose k-1} 
                    (i\tan\theta)^{2k-2}                           \cr
&=
(\cos\theta)^{u+v}\cdot i\tan\theta \,
{_2F_1}(1-u,-v;1|-\tan^2\theta)                               &(13)\cr
\noalign{\hbox{and}}
\phi_t(x) 
&=
(\cos\theta)^{u+v}\biggl[\delta(v) + \bigl(1 - \delta(v)\bigr)
\sum_{k=1}^{\infty} {u \choose k} {v-1 \choose k-1} 
                    (i\tan\theta)^{2k} \biggr]                     \cr
&=
(\cos\theta)^{u+v}\Bigl[\delta(v) - \bigl(1 - \delta(v)\bigr)
u\tan^2\theta\,{_2F_1}(1-u,1-v;2|-\tan^2\theta)\Bigr],        &(14)\cr
}
$$
where the $\delta(v)$ term counts the single $k=0$ path which 
contributes only when $v=0$, and ${_2F_1}$ is the Gauss hypergeometric 
function.  These amplitudes, together with those for an initially left 
moving path (which may be obtained from (13) and (14) by interchanging 
$u$ and $v$), define $K(t,x;0,0)$ for the quantum particle automaton.  
By linearity (additivity), therefore, they provide an exact solution 
for the evolution of any initial condition.

%--------|---------|---------|---------|---------|---------|---------|
Although Lorentz invariance is manifestly broken by the spacetime
lattice, it is regained in the continuum limit:  Let the lattice 
spacing be $\epsilon$, \ie, replace $\theta$ by $\epsilon\theta$ and 
$(u,v)$ by $(u/\epsilon,v/\epsilon)$.  Then for $u,v \not= 0$, as 
$\epsilon \to 0$,
$$
\eqalignno{
\phi_t(x-1) 
&\sim 
i\tan\epsilon\theta\,
{_0F_1}(-;1|-(uv/\epsilon^2)\tan^2\epsilon\theta)                  \cr
&\sim
i\epsilon\theta J_0(\tau\theta)                               &(15)\cr
\noalign{\hbox{and}}
\phi_t(x)
&\sim
-(u/\epsilon)\tan^2\epsilon\theta\,
{_0F_1}(-;2|-(uv/\epsilon^2)\tan^2\epsilon\theta)                  \cr
&\sim
-(u\epsilon/\tau) J_1(\tau\theta),                            &(16)\cr
}
$$
where ${_0F_1}$ is obtained by taking the confluent limit of ${_2F_1}$
twice [28], $J_i$ is the $i$th order Bessel function of the first 
kind, and $\tau := 2\sqrt{uv} = \sqrt{t^2 - x^2}$ is the spacetime 
separation of $(0,0)$ and $(t,x)$.  The limits (15) and (16), together
with their initially left moving counterparts (still obtained by 
interchanging $u$ and $v$), give exactly the continuum propagator for
the Dirac equation for a particle of mass $\theta$ [29].  When 
$\theta = 0$, the contributions from (15) and (16) vanish and the
propagator has support only on the lightcone (the $\delta(v)$ term in
(14)).  These results explain our observations about the simulations
shown in Figures 1, 2 and 3:  For $\theta = 0$ the propagation is at
speed 1 in lattice units (\ie, is along the lightcone); as $\theta$ 
increases the `mass' $\tan\theta$ increases and the speed decreases; 
when $\theta = \pi/2$,the `mass' is infinite and there is no 
propagation.

\medskip
\noindent{\bf 5.  Two component quantum cellular automata}
\nobreak

\nobreak
%--------|---------|---------|---------|---------|---------|---------|
\noindent In the chiral representation, the one dimensional Dirac
equation describes the evolution of a two component complex spinor 
[29].  Although our interests in this paper are primarily to 
investigate possible QCA rather than to construct discrete models for
fundamental physical processes [30], the results from the previous 
section indicate that in the quantum particle automaton it is natural 
to combine the amplitudes for the single particle leaving position $x$ 
to the left and to the right into a two component field 
$\psi_t(x) := \bigl(\phi_t(x-1),\phi_t(x)\bigr)$ for 
$x \equiv t$ (mod 2) and motivate consideration of 
{\sl two component\/} QCA.  In terms of $\psi$, the evolution rule 
(9) becomes
$$
\psi_{t+1}(x)
=
\pmatrix{0 & i\sin\theta \cr
         0 &  \cos\theta \cr
        }
\psi_t(x-1)
+
\pmatrix{ \cos\theta & 0 \cr
         i\sin\theta & 0 \cr
        }
\psi_t(x+1).                                                 \eqno(17)
$$
The most general local evolution rule for a two component QCA still 
has the form (3), with $\phi$ replaced by the two component field 
$\psi$ defined at {\sl all\/} cells and the coefficients $w$ now 
representing $2 \times 2$ matrices.  The evolution rule (17) for our 
quantum particle automaton already shows that the conclusion of the 
No-go Lemma can be evaded in a two component, one dimensional QCA; the 
issue becomes identifying all possible local, homogeneous, unitary 
evolution matrices $U$ as in (4).  For $r = 1$, the most general local 
evolution rule is
$$
\psi_{t+1}(x) 
= w_{-1}\psi_t(x-1) + w_0\psi_t(x) + w_{+1}\psi_t(x+1)       \eqno(18)
$$
and the unitarity constraints are still those given by equations 
($5_m$):
$$
\eqalignno{
w_{-1}^{\vphantom{\dagger}}w_{-1}^{\dagger} + 
w_0^{\vphantom{\dagger}}w_{0\phantom-}^{\dagger} + 
w_{+1}^{\vphantom{\dagger}}w_{+1}^{\dagger}         &= I
                                                         &(19_{-1})\cr
w_0^{\vphantom{\dagger}}w_{-1}^{\dagger} + 
w_{+1}^{\vphantom{\dagger}}w_{0\phantom+}^{\dagger} &= 0
                                                            &(19_0)\cr
w_{+1}^{\vphantom{\dagger}}w_{-1}^{\dagger}         &= 0,
                                                         &(19_{+1})\cr
}
$$
together with their Hermitian conjugate equations.

%--------|---------|---------|---------|---------|---------|---------|
Parity invariance imposes two additional constraints:
$$
\eqalign{
w_{-1} &= Pw_{+1}P^{-1} \cr
w_0    &= Pw_0P^{-1}    \cr
}
\qquad
\eqalign{
(P &:= \pmatrix{0 & 1 \cr
                1 & 0 \cr
               }).          \cr
}
                                                \eqno\eqalign{&(20)\cr
                                                              &(21)\cr
                                                             }
$$
Equation ($19_{+1}$) implies that at least one of $w_{-1}$ and 
$w_{+1}$ is singular; by (20) both are, have the same eigenvalues, and 
hence can be simultaneously row/column reduced to, say,
$$
w_{-1} = \pmatrix{0 & iae^{i\alpha} \cr
                  0 &  be^{i\beta}  \cr
                 }
\qquad
w_{+1} = \pmatrix{ be^{i\beta}  & 0 \cr
                  iae^{i\alpha} & 0 \cr
                 }.                                          \eqno(22)
$$
That is, any parity invariant, $r = 1$, one dimensional, two component
QCA is unitarily equivalent to one with $w_{-1}$ and $w_{+1}$ in the
form (22).  Equation (21) implies
$$
w_0 = \pmatrix{ ce^{i\gamma} & ide^{i\delta} \cr
               ide^{i\delta} &  ce^{i\gamma} \cr
              }.                                             \eqno(23)
$$
Using the forms (22) and (23) in equation ($19_0$) shows that
$\gamma = \alpha$, $\delta = \beta$, and $d = -cb/a$.  Finally, 
equation ($19_{-1}$) forces
$$
a^2 + b^2 + c^2(1 + b^2/a^2) = 1
$$
and shows that the only nonzero solution occurs when $\alpha = \beta$.
Thus we may factor out an overall phase and reparameterize to find 
that any nontrivial solution to equations ($19_m$), (20) and (21) is 
unitarily equivalent to 
$$
\setbox1=\hbox{%
$
w_{-1} = \cos\rho\pmatrix{0 & i\sin\theta \cr
                          0 &  \cos\theta \cr
                         }
\qquad
w_{+1} = \cos\rho\pmatrix{ \cos\theta & 0 \cr
                          i\sin\theta & 0 \cr
                         }
$
}
\eqalign{
w_{-1} = \cos\rho\pmatrix{0 & i\sin\theta \cr
                          0 &  \cos\theta \cr
                         }
\qquad
w_{+1} = \cos\rho\pmatrix{ \cos\theta & 0 \cr
                          i\sin\theta & 0 \cr
                         }                                         \cr
\hbox to\wd1{\hfil%
$
w_0  =   \sin\rho\pmatrix{  \sin\theta & -i\cos\theta \cr
                          -i\cos\theta &   \sin\theta \cr
                         }.
$ 
\hfil}                                                             \cr
}
                                                             \eqno(24)
$$

%--------|---------|---------|---------|---------|---------|---------|
Our quantum particle automaton, with evolution rule given by (17), is 
the most general $\rho = 0$ solution; the one dimensional lattice 
Boltzmann equation of Succi and Benzi [18] is unitarily equivalent to 
ours, as is the one dimensional version of Bialynicki-Birula's unitary 
CA for the Dirac equation [19].  As two component QCA with 
$\rho = 0$, each of these models consists of a pair of independent 
automata supported on the spacetime cells $t + x \equiv 0$ and 1 
(mod 2), respectively.  Setting $\rho \not\equiv 0$ (mod $\pi$) 
couples these two automata.  Figure 8 shows two simulations of a 
particle initially localized to be right moving from $x = 0$, evolving 
with the same value of $\theta$, but different values of $\rho$.  Cell 
darkness is (positively) proportional to the probability 
$\psi^{\dagger} \psi$, where $\psi^{\dagger}$ is the conjugate 
transpose of $\psi$.  Increasing the coupling $\rho$ towards $\pi/2$ 
has the expected effect of decreasing the propagation speed.  

%--------|---------|---------|---------|---------|---------|---------|
\pageinsert
$$
\epsfxsize=\thirdwidth\epsfbox{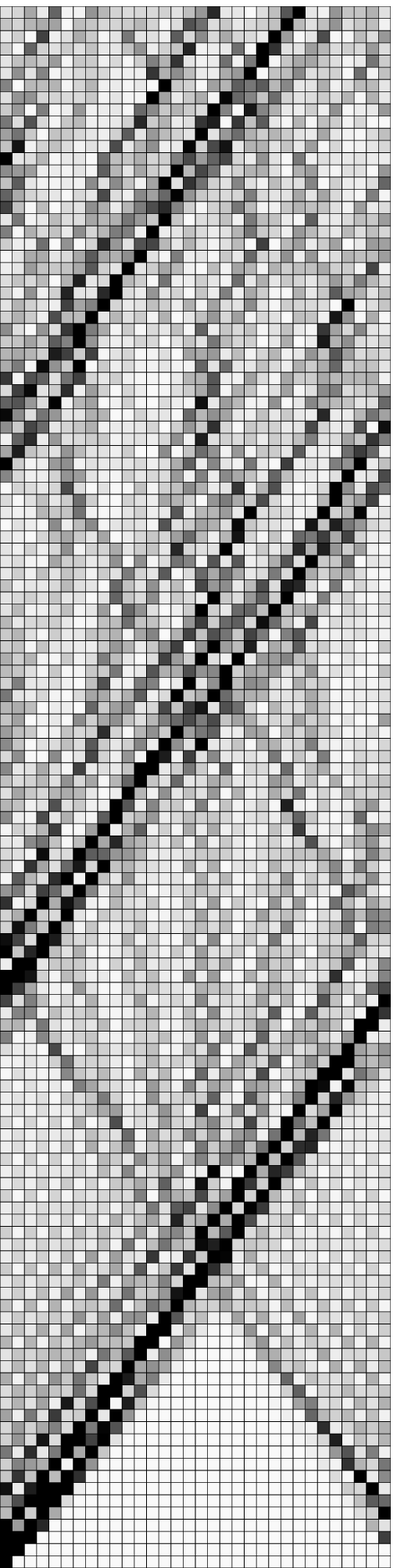}\hskip0.75truein%
\epsfxsize=\thirdwidth\epsfbox{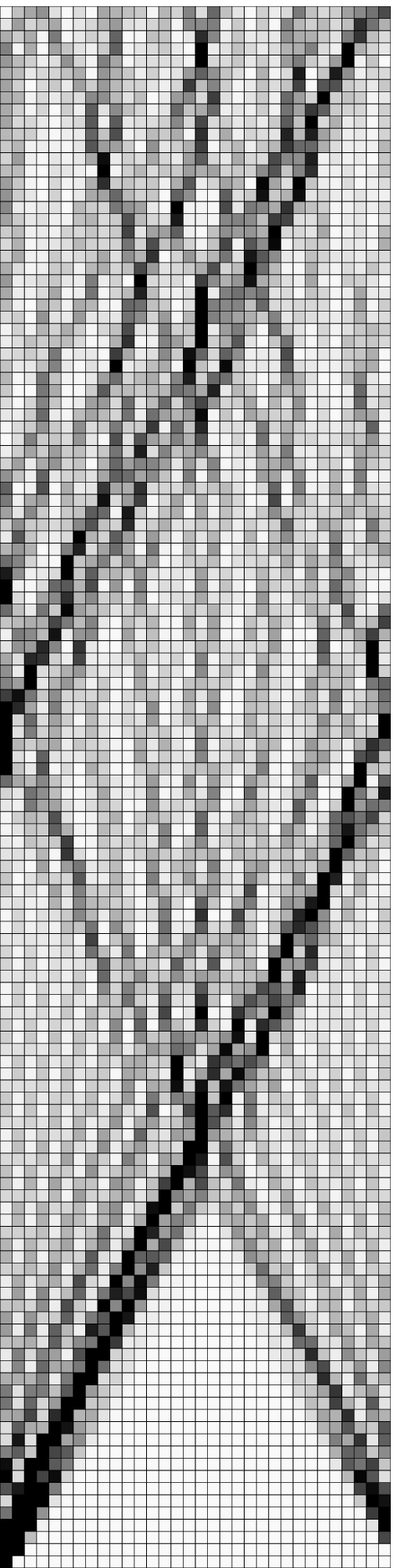}
$$
\eightpoint{%
{\narrower\noindent{\bf Figure 8}.  Two simulations of the $r = 1$ two 
component QCA starting with the same initial conditions.  In both, 
$\theta = \pi/4$, while $\rho = \pi/6$ on the left and $\pi/3$ on the 
the right.\par}
}
\endinsert

\medskip
\noindent{\bf 6.  Quantum lattice gas automata}
\nobreak

\nobreak
%--------|---------|---------|---------|---------|---------|---------|
\noindent Having reinterpreted our original QCA (9) as a quantum 
particle automaton (17), the lattice gas paradigm [17] suggests a 
different generalization than the two component QCA with evolution 
rule (18), namely to {\sl multiple\/} particles.  A LGA is a CA in 
which the possible states of the field are taken to represent 
occupation by $\alpha$-moving particles.  The evolution rule consists 
of two stages:  each particle at $x$ jumps to $x + \alpha$ and then 
the particles now at $x$ interact to possibly change their directions 
of motion.

%--------|---------|---------|---------|---------|---------|---------|
In the single particle interpretation of our QCA developed in Section 
4, the wave function $\phi_t = \sum_x \phi_t(x)|x\rangle$ describes 
the quantum state of the system at time $t$, where $|x\rangle$ is the 
one particle state defined in (10).  The scattering matrix $S$ given 
in (8) may be taken to act on the basis states $|x\rangle$ rather 
than on the coefficient amplitudes $\phi_t(x)$:
$$
S|x\rangle_t 
= \cases{ \cos\theta\,|x-1\rangle_{t+1} + 
         i\sin\theta\,|x\rangle_{t+1}   & $t\not\equiv x$ (mod 2)  \cr
         i\sin\theta\,|x\rangle_{t+1} +
          \cos\theta\,|x+1\rangle_{t+1} & $t\equiv x$ (mod 2).     \cr
        }
$$
That is, the left (right) moving particle jumps to the left (right) 
and either continues in the same direction with amplitude 
$\cos\theta$ or reverses direction with amplitude $i\sin\theta$; 
this is exactly the form of a (quantum) LGA evolution rule.

%--------|---------|---------|---------|---------|---------|---------|
Generalization to multiple particles has both kinematical and 
dynamical aspects [31]:  The Hilbert space must be extended to have
basis states $|x_1,\ldots,x_n\rangle$ (denoting the configuration with 
$n$ particles in states $x_1,\ldots,x_n$).  The familiar restriction 
in classical LGA to occupation numbers 0 and 1, \ie, an 
{\sl exclusion principle}, is consistent with the fermionic character 
of the Dirac equation we found in the macroscopic limit of the quantum
particle automaton.  With this constraint the Hilbert space has 
dimension $2^N$ (if it is $N$ for the corresponding QCA) since no two 
of the $x_i$ in a basis state may be identical.  Each particle still 
jumps at each timestep, but now there is the possibility that two 
particles (although no more than two if the exclusion principle is in 
effect) will jump to the same position at the same time.  Thus the 
scattering matrix $S$ must also be extended to include amplitudes 
$S_{ij}$ for the transitions $i \leftarrow j$ where 
$00 \le i,j \le 11$ (in binary) and the position of the 1s in an index 
indicates the occupied particle states in the pair of cells (the 
position) under consideration.  Retaining parity invariance from the 
one particle model and imposing particle number conservation, the most 
general local evolution rule is defined by the scattering matrix
$$
S = \bordermatrix{                             &
\scriptscriptstyle{\phantom{\nearrow\nwarrow}} &
\scriptscriptstyle{\phantom{\nearrow}\nwarrow} & 
\scriptscriptstyle{\nearrow\phantom{\nwarrow}} & 
\scriptscriptstyle{\nearrow\nwarrow}         \cr
\scriptscriptstyle{\phantom{\nwarrow\nearrow}} & a &   &   &   \cr
\scriptscriptstyle{\phantom{\nwarrow}\nearrow} &   & b & c &   \cr
\scriptscriptstyle{\nwarrow\phantom{\nearrow}} &   & c & b &   \cr
\scriptscriptstyle{\nwarrow\nearrow}           &   &   &   & f \cr
                 },
$$
which, just as in Section 2, must be unitary to ensure unitary 
evolution of the whole automaton.  Dividing out an overall phase, $S$
may be parameterized as
$$
S = \pmatrix{
 1 &                        &                        &             \cr
   & ie^{i\alpha}\sin\theta &  e^{i\alpha}\cos\theta &             \cr
   &  e^{i\alpha}\cos\theta & ie^{i\alpha}\sin\theta &             \cr
   &                        &                        & e^{i\beta}  \cr
}.                                                           \eqno(25)
$$
Thus individual particles evolve just as before except for a 
multiplicative phase of $e^{i\alpha}$ at each timestep, unless two 
jump to the same position at the same time, in which case they exit 
multiplied by the phase $e^{i\beta}$.  This is the simplest QLGA in
one dimension.

%--------|---------|---------|---------|---------|---------|---------|
To simulate a deterministic or probabilistic LGA, it is sufficient to
store only a single particle configuration at each timestep and then
evolve it to the next with the appropriate probability (1 in the 
deterministic case).  Each run produces a single final configuration;
multiple runs produce the same probability distribution of final 
configurations as would computing the whole Markov process, \ie, 
multiplying a vector representing state probabilities by the Markov
evolution matrix analogous to $U$.  Because probabilities do not 
add in the quantum situation---there is interference in the coherent 
sum of amplitudes---only the latter procedure is viable for a QLGA.  
For QLGA the dimension of the Hilbert space is exponential in the 
cardinality $N$ of the lattice, so simulation is potentially 
exponentially slower than for deterministic/probabilistic LGA.  With 
particle number conservation, however, the problem is not completely 
intractable:  for $n$ fermionic particles the Hilbert (or Fock, as it
would be called in the quantum field theory context) space has 
dimension ${N \choose n}$; when $n = 1$ this is just the $N$ 
amplitudes computed at each step in the one particle simulations of 
Figures 1, 2, 3 and 8.

%--------|---------|---------|---------|---------|---------|---------|
\pageinsert
$$
\epsfxsize=\thirdwidth\epsfbox{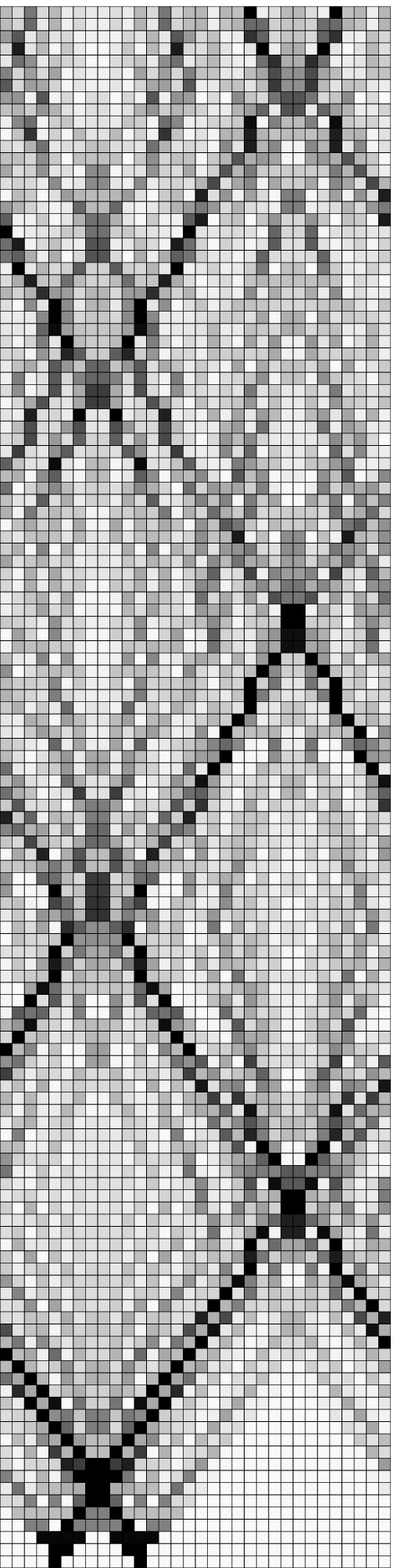}\hskip0.75truein%
\epsfxsize=\thirdwidth\epsfbox{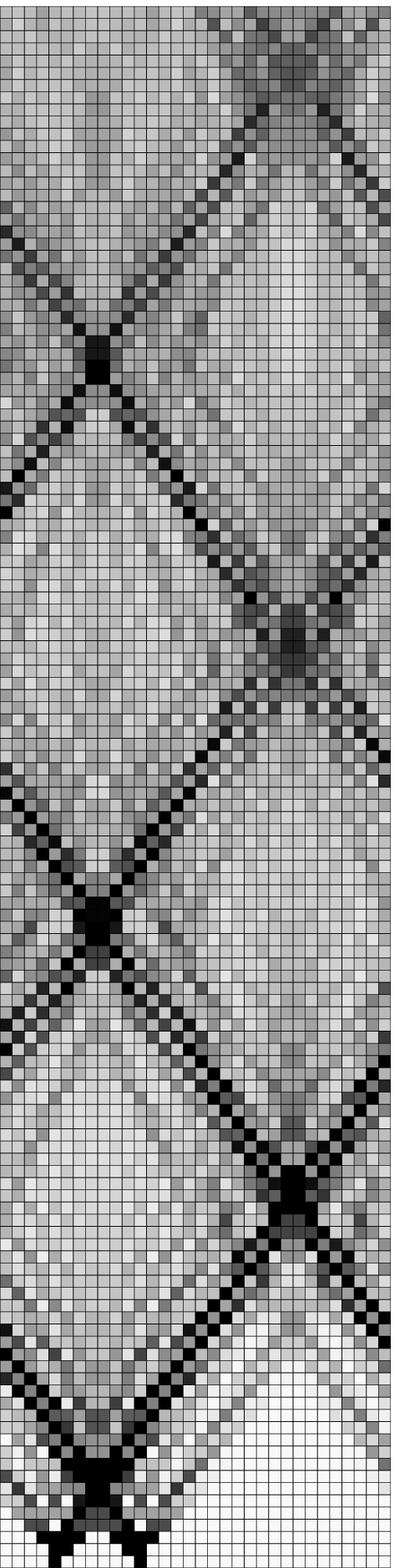}
$$
\eightpoint{%
{\narrower\noindent{\bf Figure 9}.  The QLGA with scattering matrix 
(25).  The simulation on the left contains a single particle initially 
at $x = 4$ and $x = 11$ with equal amplitude.  The one on the right 
contains two particles, initially at $x = 4$ and $x = 11$.  In both 
simulations $\theta = \pi/4$, $\alpha = 0$ and $\beta = -3\pi/4$.\par}
}
\endinsert

%--------|---------|---------|---------|---------|---------|---------|
Figure 9 shows the next simplest situation, a QLGA with $n = 2$
interacting particles.  Since one might expect that an interaction
consisting only of phase multiplication (the $e^{i\beta}$ term in $S$)
would have little effect, Figure 9 compares the two particle 
simulation with particles initially at $x = 4$ (right moving) and 
$x = 11$ (left moving) to a one particle simulation with the particle
initially at $x = 4$ and $x = 11$ with equal amplitude.  In each 
simulation cell darkness is (positively) proportional to the 
probability that a particle is present.  The qualitative difference is 
immediately apparent.

%--------|---------|---------|---------|---------|---------|---------|
The difficulty of simulating the $n$ particle sector of the Fock space
clearly grows polynomially with $n \ll N$.  But note that there is a 
duality between the particles and the `holes' (the unoccupied states)
which suggests that they should be considered to be 
{\sl antiparticles}---not surprising since we saw in Section 4 that
the one particle sector limits to the Dirac equation.  In fact, the
scattering matrix (25) is the Minkowski space form of the Boltzmann
weight in the symmetric six-vertex model, which is itself a 
specialization of the Boltzmann weight for the eight-vertex model:
$$
S = \bordermatrix{                   &
\scriptscriptstyle{\swarrow\searrow} &
\scriptscriptstyle{\swarrow\nwarrow} & 
\scriptscriptstyle{\nearrow\searrow} & 
\scriptscriptstyle{\nearrow\nwarrow}                 \cr
\scriptscriptstyle{\searrow\nearrow} & a &   &   & d \cr
\scriptscriptstyle{\searrow\nearrow} &   & b & c &   \cr
\scriptscriptstyle{\nwarrow\swarrow} &   & c & b &   \cr
\scriptscriptstyle{\nwarrow\nearrow} & d &   &   & a \cr
                 },
$$
where the downward pointing arrows represent the antiparticles (the
$CP$ duals) of the original particles.  Not only are these models
exactly solvable [32], but in the appropriate limit (critical point)
the Minkowski space symmetric six-vertex model becomes the massive
Thirring model [33].  As usual, of course, exactly solvable does not
mean that expectation values of all observables can be computed; even
in the restricted parameter domain corresponding to probabilistic
evolution only certain correlation functions have been computed [34].
This leaves a wide range of problems which may be most easily solved
by simulation with this QLGA.
\vfill\eject

%\medskip
\noindent{\bf 7.  Discussion}
\nobreak

\nobreak
%--------|---------|---------|---------|---------|---------|---------|
\noindent Motivated by the vision of quantum computation implemented 
at the device level as a QCA, we have investigated the simplest
possible models in one dimension.  Although elementary, the No-go 
Lemma and its proof seem to be original.  Evading its conclusion in 
order to construct a nontrivial QCA led us to the 
partitioning/alternating evolution rule which, taking the 
contrapositive of H\'enon's dictum,%
\sfootnote*{``Lattice gases are at present often referred to as 
            {\sl cellular automata}.  In the present note I wish to 
            advance the thesis that nothing is gained by this 
            practice, and that something is lost.'' [35].}
we interpreted as a quantum particle automaton and then generalized to 
a QLGA.  Although equivalent and similar models have been considered 
previously as regularizations of quantum field theories [18,19,33], in 
the probabilistic domain [34], and in the context of hidden variable 
theories [36], ours seems to be the first unitary simulation.

%--------|---------|---------|---------|---------|---------|---------|
Decoherence of the quantum state either in memory registers [22] or 
acted on by logical gates [23] places tight constraints on the number 
of computational steps which a quantum computer might perform 
reliably.  QCA provide a simple model in which to study decoherence 
during time evolution.  To explore the potential for quantum 
computation with QCA we are currently simulating decoherence with 
these models, as well as investigating the possibility of relaxing the 
homogeneity condition to reflect the presence of local 
gates/devices/defects, non-periodic boundary conditions, and extension 
to higher dimensions [30].

%--------|---------|---------|---------|---------|---------|---------|
These investigations are closely related to several issues in 
fundamental physics.  From the assumptions of discreteness, locality, 
unitarity, (near) homogeneity and parity invariance we were led to a 
theory of chiral fermions in one dimension.  In fact, the No-go Lemma 
was named to evoke the well known Nielsen--Ninomiya Theorem concerning 
the doubling of fermions on the lattice [37], although the logic here 
is reversed:  In $1+1$ dimensions there are several resolutions to the 
problem [38]; our one component partitioning QCA is equivalent to 
Casher and Susskind's [39]; reformulating it as a two component QCA
with {\sl two\/} independently evolving fermionic automata as in 
Section 5 resurrects the problem; and generalizing to the coupled 
evolution rule given by (18) and (24) is analogous to Wilson's 
solution [40].  In a more general context than regularizing continuum
quantum field theories, the considerations involved here in imposing 
unitarity and locality on a discrete theory are also relevant to the 
causal set program for quantum gravity [30,41].  These connections are 
not surprising:  the relation between physics and computation has only 
been made more intimate by the introduction of quantum mechanics.
\medskip

\noindent{\bf Acknowledgements}

%--------|---------|---------|---------|---------|---------|---------|
\noindent It is a pleasure to thank Gary Doolen, Ron Evans, Brosl 
Hasslacher, Melanie Quong, Jeff Rabin and Hans Sieburg for 
conversations about various aspects of this project.
\vfill\eject

\medskip
\global\setbox1=\hbox{[00]\enspace}
\parindent=\wd1

\noindent{\bf References}
\medskip

\parskip=0pt
%--------|---------|---------|---------|---------|---------|---------|
\item{[1]}
P. Benioff,
``The computer as a physical system:  a microscopic quantum mechanical
  Hamiltonian model of computers as represented by Turing machines'',
\JSP\ {\bf 22} (1980) 563--591;\hfb
R. Landauer,
``Uncertainty principle and minimal energy dissipation in the 
  computer'',
\IJTP\ {\bf 21} (1982) 283--297;\hfb
\feynman,
``Quantum mechanical computers'',
\FP\ {\bf 16} (1986) 507--531;\hfb
and references therein.

\item{[2]}
\feynman,
``Simulating physics with computers'',
\IJTP\ {\bf 21} (1982) 467--488.

\item{[3]}
\deutsch,
``Quantum theory, the Church--Turing principle and the universal
  quantum computer'',
\PRSLA\ {\bf 400} (1985) 97--117.

\item{[4]}
\deutsch\ and R. Jozsa,
``Rapid solution of problems by quantum computation'',
\PRSLA\ {\bf 439} (1992) 553--558;\hfb
A. Berthiaume and G. Brassard,
``The quantum challenge to structural complexity theory'',
in {\sl Proceedings of the 7th Structure in Complexity Theory 
Conference}, Boston, MA, 22--25 June 1992
(Los Alamitos, CA:  IEEE Computer Society Press 1992) 132--137;\hfb
E. Bernstein and U. Vazirani,
``Quantum complexity theory'',
in {\sl Proceedings of the 25th ACM Symposium on Theory of Computing},
San Diego, CA, 16--18 May 1993
(New York:  ACM Press 1993) 11--20;\hfb
D. R. Simon,
``On the power of quantum computation'',
in S. Goldwasser, ed.,
{\sl Proceedings of the 35th Symposium on Foundations of Computer 
Science}, Santa Fe, NM, 20--22 November 1994
(Los Alamitos, CA:  IEEE Computer Society Press 1994) 116--123.

\item{[5]}
P. W. Shor,
``Algorithms for quantum computation:  discrete logarithms and 
  factoring'',
in S. Goldwasser, ed.,
{\sl Proceedings of the 35th Symposium on Foundations of Computer 
Science}, Santa Fe, NM, 20--22 November 1994
(Los Alamitos, CA:  IEEE Computer Society Press 1994) 124--134.

\item{[6]}
R. L. Rivest, A. Shamir and L. Adleman,
``A method of obtaining digital signatures and public-key 
  cryptosystems'',
\CACM\ {\bf 21} (1978) 120--126.

\item{[7]}
D. P. DiVincenzo,
``Two-bit gates are universal for quantum computation'',
\PRA\ {\bf 51} (1995) 1015--1022;\hfb
J. I. Cirac and P. Zoller,
``Quantum computations with cold trapped ions'',
\PRL\ {\bf 74} (1995) 4091--4094;\hfb
A. Barenco, C. H. Bennett, R. Cleve, D. P. DiVincenzo, N. Margolus,
P. Shor, T. Sleator, J. Smolin and H. Weinfurter,
``Elementary gates for quantum computation'',
\PRA\ {\bf 52} (1995) 3457--3467;\hfb
I. L. Chuang and Y. Yamamoto,
``A simple quantum computer'',
\PRA\ {\bf 52} (1995) 3489--3496.

\item{[8]}
\teich, K. Obermeyer and G. Mahler,
``Structural basis of multistationary quantum systems.  II.  
  Effective few-particle dynamics'',
\PRB\ {\bf 37} (1988) 8111--8121.

\item{[9]}
C. S. Lent and P. D. Tougaw,
``Logical devices implemented using quantum cellular automata'',
\JAP\ {\bf 75} (1994) 1818--1825.

\item{[10]}
\teich\ and G. Mahler,
``Stochastic dynamics of individual quantum systems:  stationary
  rate equations'',
\PRA\ {\bf 45} (1992) 3300--3318;\hfb
H. K\"orner and G. Mahler,
``Optically driven quantum networks:  applications in molecular 
  electronics'',
\PRB\ {\bf 48} (1993) 2335--2346.

\item{[11]}
W. D. Hillis,
``New computer architectures and their relationship to physics or
  why computer science is no good'',
\IJTP\ {\bf 21} (1982) 255--262;\hfb
N. Margolus,
``Parallel quantum computation'',
in W. H. Zurek, ed.,
{\sl Complexity, Entropy, and the Physics of Information},
proceedings of the SFI Workshop, Santa Fe, NM, 
29 May--10 June 1989,
{\sl SFI Studies in the Sciences of Complexity} {\bf VIII}
(Redwood City, CA:  Addison-Wesley 1990) 273--287;\hfb
\brosl,
``Parallel billiards and monster systems'',
in N. Metropolis and G.-C. Rota, eds.,
{\sl A New Era in Computation}
(Cambridge:  MIT Press 1993) 53--65;\hfb
M. Biafore,
``Cellular automata for nanometer-scale computation'',
\PD\ {\bf 70}\break
(1994) 415--433;\hfb
R. Mainieri,
``Design constraints for nanometer scale quantum computers'',
preprint (1993) LA-UR 93-4333, cond-mat/9410109.

\item{[12]}
S. Ulam,
``Random processes and transformations'',
in L. M. Graves, E. Hille, P. A. Smith and O. Zariski, eds.,
{\sl Proceedings of the International Congress of Mathematicians},
Cambridge, MA, 30 August--6 September 1950
(Providence, RI:  AMS 1952) {\bf II} 264--275;\hfb
J. von Neumann,
{\sl Theory of Self-reproducing Automata},
edited and completed by A. W. Burks
(Urbana, IL:  University of Illinois Press 1966).

\item{[13]}
\gz,
``Quantum cellular automata'',
\CS\ {\bf 2}\break
(1988) 197--208.

\item{[14]}
S. Fussy, G. Gr\"ossing, H. Schwabl and A. Scrinzi,
``Nonlocal computation in quantum cellular automata'',
\PRA\ {\bf 48} (1993) 3470--3477.

\item{[15]}
K. Morita and M. Harao,
``Computation universality of one-dimensional reversible (injective)
  cellular automata'',
\TIEICEJE\ {\bf 72} (1989) 758--762.

\item{[16]}
G. V. Riazanov,
``The Feynman path integral for the Dirac equation'',
\SPJETP\ {\bf 6} (1958) 1107--1113;\hfb
\feynman\ and A. R. Hibbs,
{\sl Quantum Mechanics and Path Integrals}
(New York:  McGraw-Hill 1965) 34--36.

\item{[17]}
\hpdp,
``Time evolution of a two-dimensional model system.  I.  Invariant
  states and time correlation functions'',
\JMP\ {\bf 14} (1973) 1746--1759;\hfb
\hdpp,
``Molecular dynamics of a classical lattice gas:  transport 
  properties and time correlation functions'',
\PRA\ {\bf 13} (1976) 1949--1961;\hfb
U. Frisch, B. Hasslacher and Y. Pomeau,
``Lattice-gas automata for the Navier-Stokes equation'',
\PRL\ {\bf 56} (1986) 1505--1508.

\item{[18]}
S. Succi and R. Benzi,
``Lattice Boltzmann equation for quantum mechanics'',
\PD\ {\bf 69} (1993) 327--332;\hfb
S. Succi,
``Numerical solution of the Schroedinger equation using a quantum
  lattice Boltzmann equation'',
preprint (1993) comp-gas/9307001.

\item{[19]}
I. Bialynicki-Birula,
``Weyl, Dirac, and Maxwell equations on a lattice as unitary 
  cellular automata'',
\PRD\ {\bf 49} (1994) 6920--6927.

\item{[20]}
R. Landauer,
``Is quantum mechanics useful?'',
\PTRSLA\ {\bf 353} (1995) 367--376.

\item{[21]}
M. B. Plenio and P. L. Knight,
``Realistic lower bounds for the factorization time of large numbers
  on a quantum computer'',
preprint (1995) quant-ph/9512001;\hfb
D. Beckman, A. N. Chari, S. Devabhaktuni and J. Preskill,
``Efficient networks for quantum factoring'',
preprint (1996) CALT-68-2021, quant-ph/9602016.

\item{[22]}
W. G. Unruh, 
``Maintaining coherence in quantum computers'',
\PRA\ {\bf 51} (1995) 992--997;\hfb
G. M. Palma, K.-A. Souminen and A. Ekert,
``Quantum computers and dissipation'',
\PRSLA\ {\bf 452} (1996) 567--584.

\item{[23]}
I. L. Chuang, R. Laflamme, P. Shor and W. H. Zurek,
``Quantum computers, factoring and decoherence'',
\Sc\ {\bf 270} (1995) 1633-1635;\hfb
C. Miquel, J. P. Paz and R. Perazzo,
``Factoring in a dissipative quantum computer'',
preprint (1996) quant-ph/9601021.

\item{[24]}
H. Weyl,
{\sl The Theory of Groups and Quantum Mechanics}, 
translated from the 2nd revised German edition by H. P. Robertson
(New York:  Dover 1950).

\item{[25]}
S. Wolfram,
``Computation theory of cellular automata'',
\CMP\ {\bf 96} (1984) 15--57.

\item{[26]}
P. Ruj\'an,
``Cellular automata and statistical mechanical models'',
\JSP\ {\bf 49} (1987) 139--222;\hfb
A. Georges and P. Le Doussal,
``From equilibrium spin models to probabilistic cellular automata'',
\JSP\ {\bf 54} (1989) 1011--1064.

\item{[27]}
T. Toffoli and N. H. Margolus,
``Invertible cellular automata:  a review'',
\PD\ {\bf 45} (1990) 229--253.

\item{[28]}
Y. L. Luke,
{\sl The Special Functions and Their Approximations}, vol.\ {\bf I}
(NY:  Academic Press 1969) 49.

\item{[29]}
T. Jacobson and L. S. Schulman,
``Quantum stochastics:  the passage from a relativistic to a 
  non-relativistic path integral'',
\JPA\ {\bf 17} (1984) 375--383.

\item{[30]}
\dajm,
in preparation.

\item{[31]}
\bd,
``Lattice gases and exactly solvable models'',
\JSP\ {\bf 68} (1992) 575--590.

\item{[32]}
R. J. Baxter,
{\sl Exactly Solved Models in Statistical Mechanics}
(New York:  Academic Press 1982).

\item{[33]}
C. Destri and H. J. de Vega,
``Light-cone lattice approach to fermionic theories in 2D'',
\NPB\ {\bf 290} (1987) 363--391.

\item{[34]}
D. Kandel, E. Domany and B. Nienhuis,
``A six-vertex model as a diffusion problem:  derivation of 
  correlation functions'',
\JPA\ {\bf 23} (1990) L755--L762;\hfb
P. Orland,
``Six-vertex models as Fermi gases'',
\IJMPB\ {\bf 5} (1991) 2385--2400.

\vfill\eject
\item{[35]}
M. H\'enon,
``On the relation between lattice gases and cellular automata'',
in R. Monaco, ed.,
{\sl Discrete Kinetic Theory, Lattice Gas Dynamics and Foundations of 
     Hydrodynamics},
proceedings of the workshop, Torino, Italy, 20--24 September 1988
(Singapore:  World Scientific 1989) 160--161.

\item{[36]}
H. Hrgov\v ci\'c,
``Quantum mechanics on a space-time lattice using path integrals in a
  Minkowski metric'',
\IJTP\ {\bf 33} (1994) 745--795;\hfb
T. M. Samols,
``A stochastic model of a quantum field theory'',
\JSP\ {\bf 80} (1995) 793--809.

\item{[37]}
H. B. Nielsen and M. Ninomiya,
``A no-go theorem for regularizing chiral fermions'',
\PLB\ {\bf 105} (1981) 219--223;\hfb
and references therein.

\item{[38]}
Y. Nakawaki,
``A new choice for two-dimensional Dirac equation on a spatial 
  lattice'',
\PTP\ {\bf 61} (1979) 1855--1857;\hfb
R. Stacey,
``Eliminating lattice fermion doubling'',
\PRD\ {\bf 26} (1982) 468--472;\hfb
J. M. Rabin,
``Homology theory of lattice fermion doubling'',
\NPB\ {\bf 201} (1982) 315--332.

\item{[39]}
L. Susskind,
``Lattice fermions'',
\PRD\ {\bf 16} (1977) 3031--3039.

\item{[40]}
K. G. Wilson,
``Confinement of quarks'',
\PRD\ {\bf 10} (1974) 2445--2459.

\item{[41]}
L. Bombelli, J. Lee, D. A. Meyer and R. D. Sorkin,
``Spacetime as a causal set'', 
\PRL\ {\bf 59} (1987) 521--524;\hfb
\dajm,
``Spacetime Ising models'',
UCSD preprint (1995);\hfb
\dajm,
``Induced actions for causal sets'',
UCSD preprint (1995).

\end